\documentclass[sigconf]{acmart}

\usepackage{subcaption}
\usepackage{tcolorbox}

\AtBeginDocument{%
  }

\setcopyright{acmlicensed}
\copyrightyear{2026}
\acmYear{2026}
\acmDOI{}

\acmConference[Sensemaking @ CHI '26]
  {Sensemaking @ CHI'26}
  {2026}
  {Barcelona, Spain}

\acmISBN{}

\begin{document}

\title{Design Guidance Towards Addressing Over-Reliance on AI in Sensemaking}


\author{Yihang Zhao}
\email{yihang.zhao@kcl.ac.uk}
\orcid{0009-0009-2436-8145}
\affiliation{%
  \institution{King's College London}
  \city{London}
  \country{United Kingdom}
}

\author{Wenxin Zhang}
\email{wenxin.3.zhang@kcl.ac.uk}
\orcid{0009-0007-7226-4928}
\affiliation{%
  \institution{King's College London}
  \city{London}
  \country{United Kingdom}
}

\author{Amy Rechkemmer}
\email{amy.rechkemmer@kcl.ac.uk}
\orcid{0000-0001-7572-751X}
\affiliation{%
  \institution{King's College London}
  \city{London}
  \country{United Kingdom}
}

\author{Albert Mero\~no Pe\~nuela}
\email{albert.merono@kcl.ac.uk}
\orcid{0000-0003-4646-5842}
\affiliation{%
  \institution{King's College London}
  \city{London}
  \country{United Kingdom}
}

\author{Elena Simperl}
\email{elena.simperl@kcl.ac.uk}
\orcid{0000-0003-1722-947X}
\affiliation{%
  \institution{King's College London}
  \city{London}
  \country{United Kingdom}
}
\affiliation{%
  \institution{Technical University of Munich}
  \city{Munich}
  \country{Germany}
}

\renewcommand{\shortauthors}{Zhao et al.}

\begin{abstract}
Sensemaking in collaborative work and learning is increasingly supported by GenAI systems, however, emerging evidence suggests that poorly designed GenAI systems tend to provide explicit instruction that groups passively follow, fostering over-reliance and eroding autonomous sensemaking. Group awareness tools (GATs) address this challenge through implicit guidance: rather than instructing groups on what to do, GATs externalize observable collaboration data through visualizations that reveal differences between group members to create cognitive conflict, which triggers autonomous elaboration and discussion, thereby implicitly guiding autonomous sensemaking emergence. Drawing on an initial literature search of existing GAT systems, this paper explores the design of GenAI-augmented GATs to support autonomous sensemaking in collaborative work and learning, presenting preliminary design principles for discussion.
\end{abstract}






\maketitle

\noindent\textbf{Author's Accepted Manuscript (AAM).} 
This is the Author's Accepted Manuscript version of the article: Zhao, Y., Zhang, W., Rechkemmer, A., Meroño-Peñuela, A., \& Simperl, E. (2026). Design Guidance Towards Addressing Over-Reliance on AI in Sensemaking. Accepted for publication in \textit{Sensemaking @ CHI 2026}.

\section{Extended Abstract}

Sensemaking in collaborative work and learning is increasingly supported by GenAI systems, yet emerging evidence suggests that poorly designed GenAI systems tend to provide explicit instruction that groups passively follow, fostering over-reliance and eroding autonomous sensemaking \cite{bauer2025looking, liu2023incorporating, kim2025socially, yang2025analysing}. Explicit instruction refers to a highly structured approach that breaks down learning and working into manageable, step-by-step components, leaving little room for groups to develop their own interpretations and strategies. Group awareness tools (GATs) address this challenge through implicit guidance: rather than instructing groups on what to do, GATs externalize observable collaboration data through visualizations that reveal differences between group members to create cognitive conflict, which triggers autonomous elaboration and discussion, thereby implicitly guiding autonomous sensemaking emergence \cite{kirschner2015awareness, sangin2011facilitating, schnaubert2020combining, engelmann2009knowledge, sangin2009peer}. The design question this paper addresses is whether GenAI can be integrated into GATs in a way that provides implicit guidance for supporting autonomous sensemaking in collaborative work and learning.

To illustrate the distinction between explicit instruction and implicit guidance, consider a student group working on a collaborative project. A poorly designed GenAI system might tell the group step by step what they need to improve, what each member should do next, and how they should restructure their discussion. A GAT, by contrast, might display a radar chart showing each member's self-reported knowledge levels across different topics, making differences between members visible without comment. The group sees the variation themselves, discusses what it means, and decides how to respond. The same awareness information produces very different sensemaking outcomes depending on whether it is delivered as instruction or revealed as a pattern for the group to interpret.

Traditional GATs are constrained by their reliance on structured, quantitative collaboration data, limiting the kinds of awareness information they can surface \cite{zamecnik2022team, chang2024effects}. GenAI can analyze unstructured collaboration artifacts such as discussion transcripts and document revisions, potentially surfacing awareness information about aspects of group sensemaking that resist quantification, such as whether members build on each other's ideas or where shared understanding is weaker than groups believe \cite{breideband2025feasibility, suraworachet2025university, de2024assessing, claggett2025relational}. GenAI-augmented GATs therefore represent a promising direction for enriching sensemaking support, provided that integration preserves implicit guidance rather than introduces explicit instruction.

Drawing on an initial literature search of existing GAT systems across ACM Digital Library, IEEE Xplore, and Scopus, complemented by backward snowballing, we analyzed how GATs generate, present, and support exploration of group awareness information, asking what design considerations arise when GenAI is integrated. Three considerations emerged consistently.

The first concerns where GenAI should and should not be deployed. For awareness information derivable from structured data, such as contribution counts and participation patterns, rule-based approaches remain more appropriate \cite{zamecnik2022team, chang2024effects}. GenAI's value lies in qualitative interpretation of unstructured content: assessing whether explanations demonstrate understanding, whether members build on versus talk past each other, or whether emerging tensions reflect substantive disagreement \cite{suraworachet2025university, breideband2025feasibility}. Hybrid architectures combining rule-based processing for quantitative signals with GenAI for qualitative interpretation may therefore be more appropriate than end-to-end AI pipelines.

\begin{figure}[h]
    \centering
    \begin{subfigure}[t]{0.48\linewidth}
        \centering
        \includegraphics[width=\linewidth]{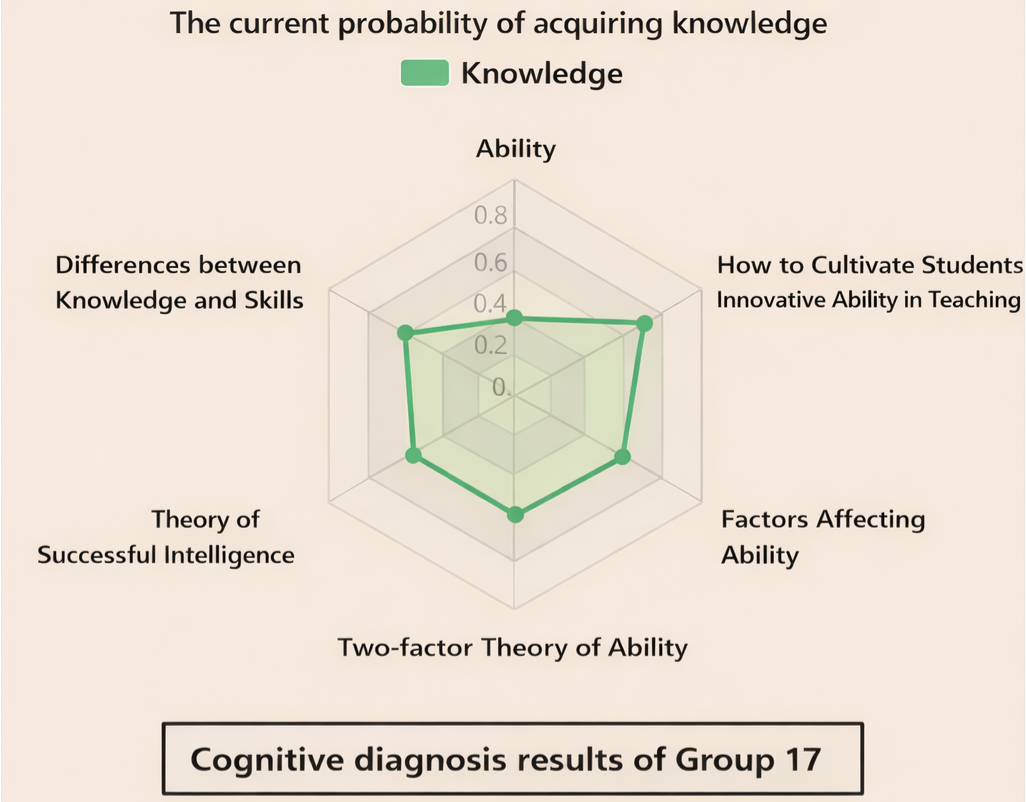}
        \caption{Without GenAI integration (reproduced from \cite{zheng2025novel}, translated into English with minor visual revisions).}
        \label{fig:RQ23}
    \end{subfigure}
    \hfill
    \begin{subfigure}[t]{0.48\linewidth}
        \centering
        \includegraphics[width=\linewidth]{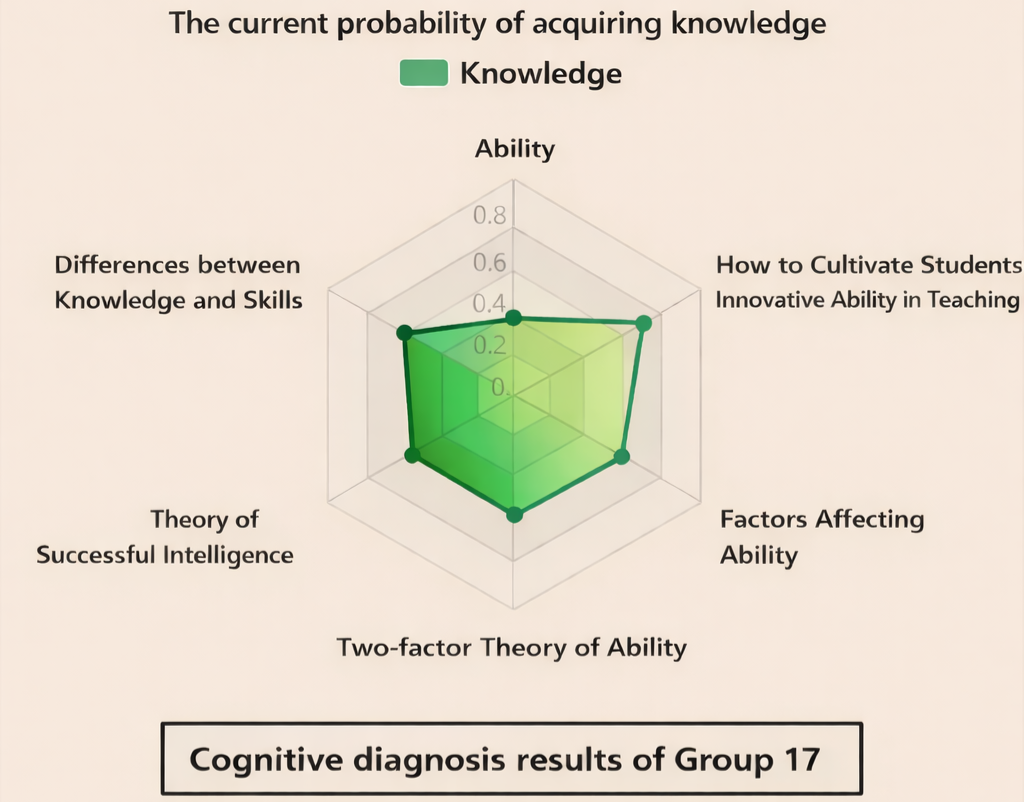}
        \caption{With GenAI integration.}
        \label{fig:RQ23G}
    \end{subfigure}
    \caption{Comparison of radar and spider chart UIs without and with GenAI integration.}
    \Description{Figure shows two radar charts side by side. The left chart displays a traditional radar chart with a polygon shape representing group self-reported knowledge levels across multiple knowledge domains. The right chart shows a GenAI-augmented version where the same polygon is preserved but axis segments have varying background colour intensities, with darker backgrounds indicating alignment between self-reported and demonstrated understanding, and lighter backgrounds indicating differences.}
    \label{fig:radar-chart-comparison}
\end{figure}

The second consideration concerns how GenAI-generated awareness information is presented. GATs reveal differences between group members to create cognitive conflict that triggers autonomous elaboration and discussion. GenAI-generated awareness information should serve this same purpose: surfacing differences that groups would otherwise be unable to see, particularly those requiring semantic understanding of unstructured content. Presenting GenAI interpretations as secondary visual encodings that augment rather than replace quantitative representations may enrich the differences revealed, creating richer cognitive conflict that triggers deeper autonomous sensemaking \cite{jarvela2023human, edwards2025human, zheng2025cognitive}. Figure~\ref{fig:radar-chart-comparison} illustrates this approach. The left chart shows a traditional radar chart of group self-reported knowledge levels. The right chart preserves this as the primary representation while encoding GenAI's independent analysis of actual discussions as background colour intensities on axis segments: darker backgrounds indicate alignment between what groups reported and what their discussions demonstrated, lighter backgrounds reveal differences. The group sees the discrepancy without being told what it means, creating cognitive conflict that may trigger autonomous elaboration and discussion.

\begin{figure}[h]
    \centering
    \includegraphics[width=0.96\linewidth]{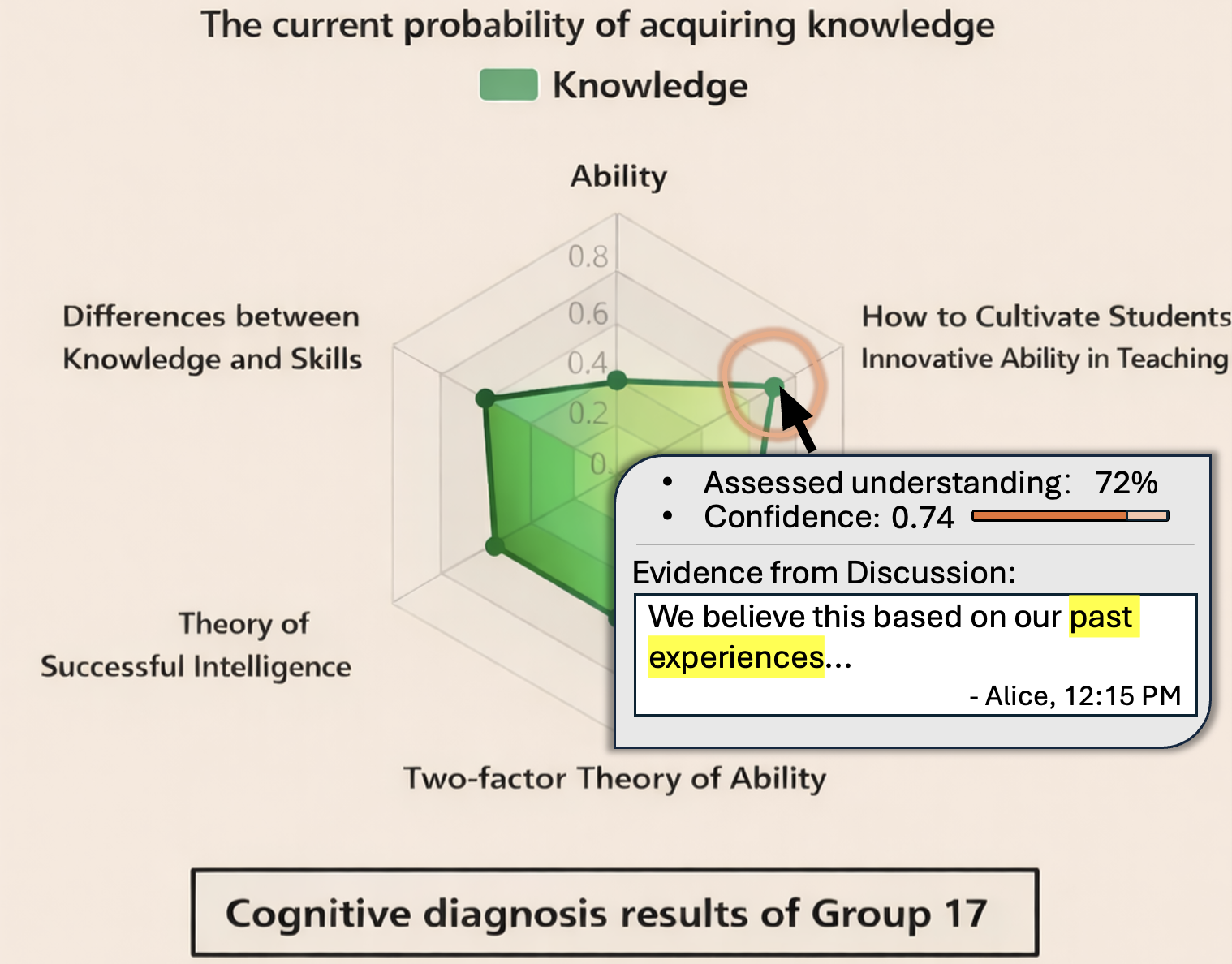}
    \caption{Hover-for-details interaction with radar charts.}
    \Description{A GenAI-augmented radar chart with background colour intensities on axis segments. A pop-up window appears when hovering over an axis, displaying GenAI's assessed understanding level, confidence score, and example discussion quotes as evidence.}
    \label{fig:interaction-hover}
\end{figure}

The third consideration concerns interaction techniques that enable groups to investigate GenAI-generated awareness information rather than passively receive it. Even when GenAI assessments are presented as secondary encodings, groups need affordances to examine the evidence underlying them in order to engage in autonomous elaboration and discussion rather than simply accepting the differences revealed. Figure~\ref{fig:interaction-hover} shows a hover-for-details interaction: hovering over a lightly shaded axis reveals GenAI's assessed understanding level, a confidence score, and example discussion quotes that served as evidence \cite{dulger2025designing, hu2025conversational, chen2021spiral, farrokhnia2025improving}. A group observing a discrepancy can read the excerpts, judge whether they agree with the interpretation, and decide whether it warrants further discussion. This positions GenAI outputs as starting points for autonomous elaboration and discussion, supporting the implicit guidance mechanism that GATs are designed to trigger.

These three considerations suggest that integrating GenAI into GATs in a way that provides implicit guidance requires attending not only to what awareness information GenAI generates, but to how it is presented and how groups are supported in engaging with it. Together, they point toward a design space in which GenAI enriches the differences revealed to groups rather than simplifying them into conclusions, preserving and potentially deepening the cognitive conflict that triggers autonomous sensemaking. Thus, we hope to discuss in this workshop what other design strategies might preserve implicit guidance in AI-supported sensemaking, and how transferable this mechanism is to sensemaking contexts beyond collaborative work and learning. Our prior research contributes expertise spanning machine learning, AI-based systems supporting individual and collaborative knowledge and creative work, interaction design with AI, and cognitive and metacognitive augmentation \cite{zhao2026ontoscope,zhao2025leveraging,zhaosurvey,zhao2024improving,zhao2024using,zhao2026exploring,zhao2024user,hu2025designing,lisena2026data,simperl2025automatic,simperl2025introducing,zhang2025trustworthy,zhao2021discriminative,luo2021survey}.


\begin{acks}
We thank Advait Sarkar from Microsoft Research Cambridge for feedback. We also acknowledge funding support, including the Engineering and Physical Sciences Research Council [grant number EP/Y009800/1], funded through Responsible AI UK (KP0011); SIEMENS AG; and the Institute for Advanced Study, Technical University of Munich, Germany.
\end{acks}

\bibliographystyle{ACM-Reference-Format}
\bibliography{SAA}





\end{document}